\documentclass[conference]{IEEEtran}
\IEEEoverridecommandlockouts

\usepackage{fancyhdr}
\usepackage{amsmath}
\usepackage{amssymb}
\usepackage{graphicx}
\usepackage{booktabs}
\usepackage{amsfonts}
\usepackage{multirow}
\usepackage{algorithm}
\usepackage{algorithmic}
\usepackage{mathrsfs}
\usepackage{bm}
\usepackage{exscale}
\usepackage{relsize}
\usepackage{setspace}
\usepackage{color}
\usepackage{comment}
\usepackage{ulem}
\usepackage{gensymb}
\usepackage{cite}

\usepackage{geometry}
\newtheorem{lemma}{\textbf{Lemma}}
\newtheorem{theorem}{\textbf{Theorem}}

\ifCLASSINFOpdf
\else
\fi

\begin{document}
	\columnsep 0.2in

\title{Sparse Estimation for XL-MIMO with Unified LoS/NLoS Representation}

\author{
	Xu~Shi\textsuperscript{1},~Xuehan~Wang\textsuperscript{1},~Jingbo~Tan\textsuperscript{1}\thanks{This work was supported by the National Postdoctoral Researcher Funding Program under Grant GZC20231376.},~Jintao~Wang\textsuperscript{1,2}\\
	\IEEEauthorblockA{
		\textsuperscript{1}Beijing National Research Center for Information Science and Technology (BNRist),\\
		Dept. of Electronic Engineering, Tsinghua University, Beijing, China\\
		\textsuperscript{2}Research Institute of Tsinghua University in Shenzhen, Shenzhen, China\\
		\{shi-x19@mails., wang-xh21@mails., tanjingbo@, wangjintao@\}tsinghua.edu.cn}
}


\maketitle

\begin{abstract}
Extremely large-scale antenna array (ELAA) is promising as one of the key ingredients for the sixth generation (6G) of wireless communications. The electromagnetic propagation of spherical wavefronts introduces an additional distance-dependent dimension beyond conventional beamspace. In this paper, we first present one concise closed-form channel formulation for extremely large-scale multiple-input multiple-output (XL-MIMO). All line-of-sight (LoS) and non-line-of-sight (NLoS) paths, far-field and near-field scenarios, and XL-MIMO and XL-MISO channels are unified under the framework, where additional Vandermonde windowing matrix is exclusively considered for LoS path. Under this framework, we further propose one low-complexity unified LoS/NLoS orthogonal matching pursuit (XL-UOMP) algorithm for XL-MIMO channel estimation. The simulation results demonstrate the superiority of the proposed algorithm on both estimation accuracy and pilot consumption. 
\end{abstract}


\IEEEpeerreviewmaketitle

\section{Introduction}

Extremely large-scale antenna array (ELAA) is a significant feature of several key candidate technologies in sixth generation (6G) mobile networks, especially for Terahertz (THz) communications \cite{XLMIMO_1,XLMIMO_2} and reconfigurable intelligent surface (RIS) \cite{gaofeifei}. Benefitted from large aperture and extremely large-scale multiple-input multiple-output (XL-MIMO) capabilities, ELAAs can greatly enhance transmission rates and network coverage. Meanwhile, spherical-wavefront electromagnetic propagation must not be overlooked for accurate beamfocusing instead of planar-wavefront \cite{nearfield_1,nearfield_2}. In contrast to the conventional beamspace (angular-domain) representation, additional distance-dependent dimension should be considered in ELAA system, which presents both challenges and promising opportunities in the coming 6G mobile networks. 

The near-field channel state information (CSI) acquisition is one of the most crucial challenges for ELAA signal processing, due to the excessive overhead and huge complexity. The near-field communications can be broadly divided into two categories, i.e., the XL-MISO and   XL-MIMO systems. The former is relatively straightforward since the 1-D channel vector is always rank-one and is easily formulated by near-field steering vector. For XL-MIMO model, the channel turns much more complex. On the one hand, the line-of-sight (LoS) path contains multiple ranks, known as intra-path multiplexing \cite{intra_path}. The product of steering vectors at transmitter (Tx)/ receiver (Rx) sides cannot be adopted directly for LoS representation. Additional Tx/Rx-intertwined coefficient should be considered for accurate channel representation. On the other hand, the non-line-of-sight (NLoS) paths show different representation from LoS one \cite{luyu}, which further deteriorates the accuracy and complexity of XL-MIMO channel formulation and estimation. 

The existing near-field CSI acquisition schemes can be summarized as follows. \cite{subarray_jinshi} first studied the ELAA and proposed one subarray-based estimation scheme. Then \cite{polardomain_estimation} introduced the polar-domain representation estimation method via compressive sensing. \cite{add_review} proposes two-phase estimation to decrease overhead and computational complexity. The chirp-based hierarchical codebook-based near-field beam training was further provided in our previous work \cite{chirp_based} for overhead reduction. Nevertheless, those above methods are merely designed for single-antenna user scenario and are not applicable for XL-MIMO scenario. To our best knowledge, limited studies have been conducted for CSI acquisition. Only \cite{sunshu} presented a hybrid spherical-planar-wave channel model, where several subdivided ULA fractions are individually considered with redundant overhead and complexity. Besides, \cite{luyu} quantized the entire spatial space and proposed an maximum-likelihood-based traversal method which suffers from the same issue.

In this paper, we aim at addressing the near-field modeling mismatch and overhead consumption problems. Most importantly, we formulate the XL-MIMO channel through a concise closed-form expression, with quite small approximation error. The distinctions between LoS and NLoS paths, spherical and planar-wavefront propagations, single and multiple-antenna user scenarios are all unified under this framework.  Furthermore, we design one low-complexity compressive sensing-based scheme called unified LoS/NLoS orthogonal matching pursuit (XL-UOMP) algorithm. Simulation results demonstrate the superiority of the proposed algorithm on both estimation accuracy and pilot consumption. 

%

\section{System Model}
\begin{figure}[!t]
	\centering
	\includegraphics[width=0.8\linewidth]{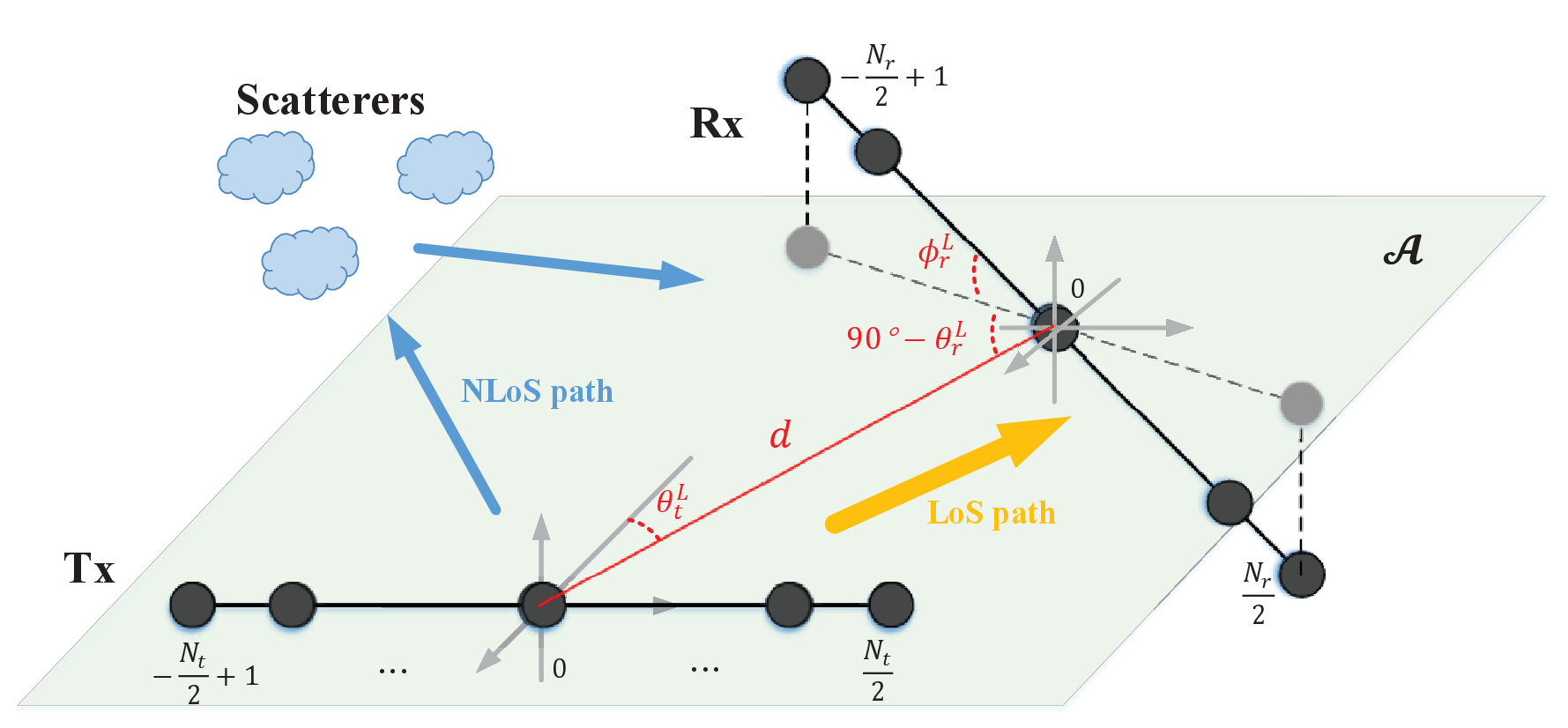}
	\caption{Block diagram of near-field XL-MIMO system model.}
	\label{system_model}
\end{figure}

We consider a narrowband point-to-point near-field mmWave system where both the transmitter and receiver are equipped with uniform linear array (ULA), as shown in Fig. \ref{system_model}. At the Tx side, $N_\text{RF}$ radio frequency (RF) chains are assigned to $N_\text{T}$ antennas via fully-connected phase shift network ($N_\text{RF}\ll N_\text{T}$), and similarly $N_\text{R}$ antennas are deployed at Rx side with $N_\text{RF}$ RF chains ($N_\text{RF}\ll N_\text{R}$). The antenna arrays at Tx/Rx sides are indexed as $n_1\in \{-\frac{N_\text{T}}{2}+1,\dots,\frac{N_\text{T}}{2}\}$ and $n_2\in \{-\frac{N_\text{R}}{2}+1,\dots,\frac{N_\text{R}}{2}\}$ from left to right, respectively. The antenna spacing is defined as half of the carrier wavelength $d=\lambda/2$. The transmission distance between the Tx/Rx antenna elements is marked as $r_{n_1,n_2}$, and particularly, $r_{0}$ denotes the transmission distance between the central elements of Tx/Rx arrays. The fully-connected precoding/combining structures are deployed at Tx/Rx sides, marked as $\bm F_t\in \mathbb{C}^{N_\text{T}\times N_\text{RF}}$ and $\bm W_t\in \mathbb{C}^{N_\text{R}\times N_\text{RF}}$ inside the $t$-th timeslot. 

We herein consider non-coplanar and non-parallel ULA Tx/Rx arrays, the formulation of which will be much more complicated and general compared with previous state-of-the-art works \cite{sunshu,luyu}. As shown in Fig. \ref{system_model}, the Tx array is set inside the plane $\mathcal{A}$ (green region spanned by Tx array vector and LoS path to Rx central point $n_2=0$ in Fig. \ref{system_model}), and Rx array is slanted outside this plane $\mathcal{A}$ with tilt angle $\phi_\text{rx}^L$. $\theta_\text{tx}^L$ denotes the angle of departure at Tx side and $\theta_\text{rx}^L$ represents the angle of arrival for the virtual Rx array projected inside plane $\mathcal{A}$ (dashed array in Fig. \ref{system_model}). The distance-dependent LoS path can be strictly formulated as

\begin{equation}
\bar{\bm H}^\text{L}(n_1,n_2)=g^\text{L}\cdot \bigg[\frac{1}{r_{n_1,n_2}} \cdot \text{exp}(-j2\pi r_{n_1,n_2}/\lambda)\bigg],
\label{LoS_init}
\end{equation}
where $\frac{1}{r_{n_1,n_2}}g^\text{L}$ denotes the distance-dependent LoS path loss. Similarly, let $r^\text{tx}_{n_1,l}$ and $r^\text{rx}_{n_2,l}$ denote the transmission distance between Tx/Rx array element and the $l$-th scatterer, respectively. Then the NLoS channel is yielded as
\begin{equation}
    \arraycolsep=0.5pt\def\arraystretch{1.5}
	\begin{array}{lll}
\bar{\bm H}^\text{NL}_l &= &g_l^\text{NL}\cdot \bigg[ \frac{1}{r^\text{tx}_{n_1,l}}\frac{1}{r^\text{rx}_{n_2,l}} \cdot \text{exp}\big(-j2\pi(r^\text{tx}_{n_1,l}+r^\text{rx}_{n_2,l})/\lambda\big)  \bigg]_{N_\text{R}\times N_\text{T}}\\
&& \ l=1\dots,L.
\end{array}
\label{NLoS_init}
\end{equation}
Therefore, the received signal at Rx side can be formulated as follows:
\begin{equation}
	\bm Y_t = \bm W_t^H \bigg(\bar{\bm H}^\text{L}+\sum_{l=1}^L \bar{\bm H}^\text{NL}_l \bigg) \bm F_t \bm s_t +\bm n_t,
\end{equation}
where $\bm n_t$ denotes the additive white Gaussian noise with variance $\sigma^2$ following $\bm n_t\sim \mathcal{CN}(0,\sigma^2\bm I_{N_\text{RF}})$. Let $\bar{\bm H}=\bar{\bm H}^\text{L}+\sum_{l=1}^L \bar{\bm H}^\text{NL}_l $ represent the total channel response with both LoS and NLoS components.

\section{LoS/NLoS Channel Approximation and Analysis}

In fact, the transmission distance ($r_{n_1,n_2}$, $r^\text{tx}_{n_1,l}$ or $r^\text{rx}_{n_2,l}$) does not varies nonlinearly with indices $n_1$ and $n_2$ that rapidly. However, under the high-frequency such as mmWave or sub-THz scenarios, the tiny wavelength $\lambda$ (almost $10^{-5}\sim10^{-3}$ metres) will provide a huge amplification effect for the nonlinear variation inside phase components (\ref{LoS_init}), (\ref{NLoS_init}). This inspires us that for each path, we can simplify the channel amplitudes into spatial-independent factors and focus more on the phase variations. 

For each NLoS path, the transmission distances between the scatterer and Tx/Rx are approximated via Taylor expansion and the former three terms are reserved under spherical wavefront assumption. The derivation is omitted for brevity since the process is quite similar to that in near-field MISO system, which has been extensively studied in \cite{polardomain_estimation,chirp_based}. Given the distance-isolated term outside the path loss, i.e., $g^\text{NL}_l$, the final approximated NLoS channel is written as follows
\begin{equation}
	\bm H^\text{NL}_l = \frac{g^\text{NL}_l e^{-j2\pi (r^\text{tx}_{0,l}+r^\text{rx}_{0,l})/\lambda}}{r^\text{tx}_{0,l}r^\text{rx}_{0,l}}\cdot\bm b_{N_\text{R}}(b^\text{rx}_l,k^\text{rx}_l)\bm b_{N_\text{T}}^H(b^\text{tx}_l,k^\text{tx}_l),
	\label{NLoS_model}
\end{equation}
where
\begin{equation}
	\bm b_N(b,k)=\big[\text{exp}(-j \pi (bn+kn^2) )\big]_{N\times 1}, \small{n=-\frac{N}{2}+1,\dots,\frac{N}{2}}
	\label{basic_steering_near_field}
\end{equation}
is the near-field steering vector. $b$ and $k$ here denote the first-order and second-order Taylor coefficient, respectively. Notice that for each NLoS path, $\bm H^\text{NL}_l$ is still rank-one matrix and sparse in polar-domain representation, which is consistent to the previous studies.

\newcounter{mytempeqncnt}
\begin{figure*}[!t]
	\normalsize
	\setcounter{mytempeqncnt}{\value{equation}}
	\setcounter{equation}{5}
	\begin{small}
	\begin{equation}
		\text{Rx}\  n_2: \ \ \bigg(r_{0}\sin \theta_\text{tx}^L, r_{0}\cos \theta_\text{tx}^L, 0 \bigg) + \bigg(d n_2 \sin(\frac{\pi}{2}-\theta_\text{rx}^L+\theta_\text{tx}^L)\cos\phi_\text{rx}^L ,d n_2 \cos(\frac{\pi}{2}-\theta_\text{rx}^L+\theta_\text{tx}^L) \cos\phi_\text{rx}^L, d n_2 \sin\phi_\text{rx}^L\bigg)
		\label{Rx_corr}
	\end{equation}
	\end{small}
	\setcounter{equation}{\value{mytempeqncnt}}
	\hrulefill
	\vspace*{4pt}
\end{figure*}
\setcounter{equation}{6}

\begin{figure*}[!t]
	\normalsize
	\setcounter{mytempeqncnt}{\value{equation}}
	\setcounter{equation}{6}
	\begin{small}
	\begin{equation}
		\arraycolsep=1.2pt\def\arraystretch{1.4}
		\begin{array}{cll}
			r_{n_1,n_2}&=&\sqrt{\bigg(r_0\sin\theta_\text{tx}^L+dn_2\sin(\frac{\pi}{2}-\theta_\text{rx}^L+\theta_\text{tx}^L)\cos\phi_\text{rx}^L-dn_1\bigg)^2 + \bigg(r_0\cos\theta_\text{tx}^L+dn_2\cos(\frac{\pi}{2}-\theta_\text{rx}^L+\theta_\text{tx}^L)\cos\phi_\text{rx}^L\bigg)^2+
				\bigg( d n_2\sin\phi_\text{rx}^L \bigg)^2}\\
			&=&\sqrt{r_0^2+\bigg(2r_0\sin\theta_\text{rx}^L\cos\phi_\text{rx}^L\cdot dn_2-2r_0\sin\theta_\text{tx}^L\cdot dn_1\bigg)+\bigg(d^2n_1^2+d^2n_2^2 -2\cos(\theta_\text{rx}^L-\theta_\text{tx}^L)\cos\phi_\text{rx}^L \cdot d^2n_1n_2\bigg)}\\
			&\overset{(a)}{\approx}&r_0+
			\underbrace{\bigg(-dn_1\sin\theta_\text{tx}^L  +  dn_2\sin \theta_\text{rx}^L\cos \phi_\text{rx}^L\bigg)}_{\text{first-order: far-field approximation}} + 
			\underbrace{\bigg(\frac{\cos^2\theta_\text{tx}^L}{r_0} d^2n_1^2 + \frac{1-\sin^2\theta_\text{rx}^L \cos^2\phi_\text{rx}^L}{r_0}d^2n_2^2 - \frac{\cos\theta_\text{tx}^L\cos\theta_\text{rx}^L\cos\phi_\text{rx}^L}{r_0}d^2n_1 n_2   \bigg)}_{\text{second-order: near-field approximation}}\\
			&=&r_0+
			\underbrace{\bigg(-dn_1\sin\theta_\text{tx}^L  +\frac{\cos^2\theta_\text{tx}^L}{r_0} d^2n_1^2 \bigg)}_{\text{Tx-dependent}} +
			\underbrace{\bigg( dn_2\sin \theta_\text{rx}^L\cos \phi_\text{rx}^L + \frac{1-\sin^2\theta_\text{rx}^L \cos^2\phi_\text{rx}^L}{r_0}d^2n_2^2 \bigg)}_{\text{Rx-dependent}}
			-\underbrace{ \frac{\cos\theta_\text{tx}^L\cos\theta_\text{rx}^L\cos\phi_\text{rx}^L}{r_0}d^2n_1 n_2 }_{\text{Tx/Rx-coupled term}}
		\end{array}
		\label{distance_formulation}
	\end{equation}
\end{small}
	\setcounter{equation}{\value{mytempeqncnt}}
	\hrulefill
	\vspace*{4pt}
\end{figure*}
\setcounter{equation}{7}

However, the near-field \emph{LoS} channel exhibits entirely different characteristics. First, the coordinates of the Tx array elements are easily obtained as $(dn_1,0,0)$, and similarly, the Rx coordinates can be yielded to (\ref{Rx_corr}). Then the transmission distance can be strictly formulated as (\ref{distance_formulation}). Through Taylor expansion in $(a)$, we approximate the  near-field  distance formulation and obtain the final second-order representation \footnote{It is reasonable to adopt Taylor expansion since the first-order is exactly the far-field transmission distance approximation, where the second-order expansion is compatible and more accurate}. In contrast to the previous near-field NLoS paths, additional Tx/Rx-intertwined second-order term  exists in the   XL-MIMO LoS channel, which limits formulation accuracy and  beamfocusing performance, or positively speaking, promotes the possibility of intra-path multiplexing \cite{intra_path}. Due to the spherical wavefront at double sides (Tx and Rx) in the near-field communication, conventional beamspace or recent polar-domain representation (corresponding to near-field \emph{MISO} scenario specifically) will suffer severe modeling distortion and beamforming degradation due to the Tx/Rx intertwined component in (\ref{distance_formulation}). How to establish the closed-form channel expression \emph{mathematically and succinctly with joint unification of LoS/NLoS styles} is significant and challenging, which will greatly benefit CSI acquirement and beamfocusing design.  

Substituting (\ref{distance_formulation}) into the LoS channel, we obtain the approximated closed-form   LoS representation:
\begin{equation}
	\bm H^\text{L}=\frac{g^\text{L}e^{-j2\pi r_0/\lambda}}{r_0}\cdot\bigg(\bm b_{N_\text{R}}(b^\text{rx}_0,k^\text{rx}_0)\bm b^H_{N_\text{T}}(b^\text{tx}_0,k^\text{tx}_0)\bigg)\odot \bm V_0,
	\label{LoS_model}
\end{equation}
where
\begin{equation}
	\bm V_0=\bigg[ \text{exp}(j2\pi\cdot \omega_0 \cdot n_1 n_2) \bigg]_{N_\text{R}\times N_\text{T}}
	\label{Van_LoS}
\end{equation}
is a specific Vandermonde windowing matrix that acts on the conventional steering matrix $\bm b_{N_\text{R}}\bm b^H_{N_\text{T}}$ via element-wise product. The detailed parameters in (\ref{LoS_model}) are formulated as follows:
\begin{equation}
	\left\{
\arraycolsep=1.0pt\def\arraystretch{1.8}
\begin{array}{cll}
		b^\text{rx}_0&=&\displaystyle\sin \theta_\text{rx}^\text{L} \cos \phi_\text{rx}^\text{L}\\
		k^\text{rx}_0&=&\displaystyle\frac{\lambda(1-\sin^2\theta_\text{rx}^L \cos^2\phi_\text{rx}^L)}{2r_0}\\
		b^\text{tx}_0&=&\displaystyle\sin \theta_\text{tx}^\text{L}\\
		k^\text{tx}_0&=&\displaystyle -\frac{\lambda\cdot\cos^2\theta_\text{tx}^L}{2r_0} \\
		\omega_0&=&\displaystyle \frac{\lambda\cdot\cos\theta_\text{tx}^L\cos\theta_\text{rx}^L\cos\phi_\text{rx}^L}{4r_0}
\end{array},
\label{detailed_para}
\right.
\end{equation}
where $\{b^\text{rx}_0,k^\text{rx}_0\}$ represents the beamforming direction at Rx side while $\{b^\text{tx}_0,k^\text{tx}_0\}$ represents the beamforming direction at Tx side. More importantly, $\omega_0$ here controls the degree of freedom (DoF) of channel matrix. More DoFs (or ranks) can be obtained with increased value of $\omega_0$, which also means larger channel approximation error via conventional polar-domain representation. Instead of estimating the spatial four-parameter group $\{\theta_\text{tx}^\text{L}, \theta_\text{rx}^\text{L},\phi_\text{rx}^\text{L},r_0\}$, we herein consider the rapid compressive-sensing-based estimation of the five-element tuple in (\ref{detailed_para}). The concise expression and independent relationships  here can greatly simplify the estimation process, reduce complexity, and minimize overhead. And the following theorem for LoS channel can be yielded :
\begin{theorem}
	The LoS channel  $\bm H^\text{L}$  contains full rank $\min \{N_\text{T},N_\text{R}\}$ mathematically, but the number of \emph{available} pipelines here can be approximated as $\omega_0N_\text{T}N_\text{R}$.
	\label{theorem1}
\end{theorem}
\begin{IEEEproof}
	See Appendix.
\end{IEEEproof}

Specifically, when we set the windowing parameter $\omega_0=0$, the windowing Vandermonde matrix turns into $\bm V=\bm 1_{N_\text{R}\times N_\text{T}}$. Then the LoS channel is further simplified to rank-one matrix, which contains the same pattern with NLoS paths (\ref{NLoS_model}). On the contrary, when we set the windowing parameter $\omega_0=1$, the Vandermonde matrix ($1\sim \min\{N_\text{T},N_\text{R}\}$ columns and rows) will turn into discrete Fourier transform (DFT) matrix, with full rank $\min\{N_\text{T},N_\text{R}\}$. In   near-field communications, the windowing parameter should follow $0\leq \omega_0\ll \frac{1}{\min\{N_\text{T},N_\text{R}\}}$ in practice. Therefore, it is easily observed that the NLoS channel (\ref{NLoS_model}) is one specific case of LoS representation with $\omega_0=0$ and thus we can gather them into one unified mathematical framework.

For convenience, we omit the inner parameter in LoS/NLoS steering vector and rewrite them as $\bm b_{N_\text{R},0},\bm b_{N_\text{T},0}$ and $\bm b_{N_\text{R},l},\bm b_{N_\text{T},l} , l=1,\dots,L$, respectively. Besides, we define the total LoS and NLoS path loss as $\beta_0$ and $\beta_l,l=1,\dots,L$, respectively.
Then the unified closed-form expression of noiseless received signal is reformulated as 
\begin{equation}
	\arraycolsep=1.0pt\def\arraystretch{1.2}
	\begin{small}
\begin{array}{cll}
	\bm y_t &=&\displaystyle \bm W_t^H \cdot \bigg(\beta_0(\bm b_{N_\text{R},0}\bm b_{N_\text{T},0}^H)\odot \bm V +\sum_{l=1}^L\beta_l\bm b_{N_\text{R},l}\bm b_{N_\text{T},l}^H \bigg)\cdot  \bm F_t \bm s_t \\
	&=&\displaystyle \bm W_t^H \cdot \bigg(\sum_{l=0}^L \beta_l\cdot (\bm b_{N_\text{R},l}\bm b_{N_\text{T},l}^H)\odot \bm V_l \bigg)\cdot  \bm F_t \bm s_t ,
	\end{array}
\end{small}
	\label{unified_received_signal}
\end{equation}
where the $0$-th path denotes the LoS one while the rest $L$ paths represent the NLoS ones. $\bm V_l$ is the Vandermonde windowing matrix following the same pattern in (\ref{Van_LoS}) with inner parameter $\omega_l$. For all NLoS paths, to guarantee the rank-one matrix characteristic, $\omega_l$ should always keep zero except the LoS case $l=0$.

%

\section{Unified LoS/NLoS Sparse Estimation}

Under the unified framework about received signal in (\ref{unified_received_signal}), the critical point is the dictionary matrix establishment for compressive sensing. First we make vectorization for the noiseless received signal as
\begin{equation}
\arraycolsep=1.0pt\def\arraystretch{1.2}
\begin{small}
\begin{array}{cll}
	\bm y_t &=& \displaystyle\bigg((\bm F_t\bm s_t)^T \otimes \bm W_t^H\bigg) \cdot \text{vec}\bigg(\sum_{l=0}^L \beta_l\cdot (\bm b_{N_\text{R},l}\bm b_{N_\text{T},l}^H)\odot \bm V_l\bigg)\\
	& = & \displaystyle  \bigg((\bm F_t\bm s_t)^T \otimes \bm W_t^H\bigg) \cdot \bigg(\sum_{l=0}^L \beta_l\cdot (\bm b_{N_\text{T},l}^*\otimes \bm b_{N_\text{R},l} \odot \bm v_l)\bigg)\\
	&=&\displaystyle  \bigg((\bm F_t\bm s_t)^T \otimes \bm W_t^H\bigg) \cdot \bm G \cdot \bm \beta,
\end{array}
\end{small}
\label{formula_OMP}
\end{equation}
where $\bm v_l=\text{vec}(\bm V_l)$ is the vectorization of windowing matrix $\bm V_l$, $\bm\beta=[\beta_0,\dots,\beta_L]^T$ and $\bm G=[\bm b_{N_\text{T},l}^*\otimes \bm b_{N_\text{R},l} \odot \bm v_l],l=0,\dots,L$ represent the path gain vector and generalized steering matrix, respectively. Without loss of generality, we define $\bm \psi_t = (\bm F_t\bm s_t)^T \otimes \bm W_t^H$ as the sensing vector and collect multiple pilot slots $t=1,\dots, T$, i.e., $\bm \Psi=[\bm \psi_1^T,\dots, \bm \psi_T^T]^T$. Then the total received signal can be formulated as $\bm y = \bm \Psi \bm G\bm \beta$ with the total channel $\text{vec}(\bm H)=\bm G\bm \beta$, where $\bm y=[\bm y_1^T,\dots,\bm y_T^T]^T$.

Inspired by the far-field beamspace and polar-domain representation, we can similarly characterize the sparsity for the   XL-MIMO system. Inside far-field beamspace representation \cite{far_field_OMP}, only the angular directions, i.e., $b^\text{rx}, b^\text{tx}$ in (\ref{detailed_para}), are quantized to form several spatial-stationary array steering vectors for sparse representation. Inside polar-domain representation,  additional distance-related parameters, i.e., $k^\text{rx}, k^\text{tx}$ in (\ref{detailed_para}), are
quantized to generate spatial-nonstationary beams. Due to the further Tx/Rx coupled windowing parameter $\omega_0$ here, we have to quantize this parameter and generate a huge codebook for the sparse channel representation. Assume the uniformly quantized values are marked as $\{b_q^\text{tx/rx}\}, \{k_q^\text{tx/rx}\}$ and $\{\omega_{0,q}\}$ and the total quantized numbers are $N_b^\text{tx/rx}$, $N_k^\text{tx/rx}$ and $N_{\omega}$, respectively. Thus we can generate each column of dictionary matrix via the five parameters and obtain the huge representation matrix and corresponding channel as $\hat{\bm G}\in \mathbb{C}^ {N_\text{T}N_\text{R}\times N_b^\text{tx}N_b^\text{rx}N_k^\text{tx}N_k^\text{rx}N_{\omega}}$ and $\text{H}=\hat{\bm G}\hat{\bm \beta}$, respectively, where $\hat{\bm G}$ is predefined with each column generated by one quantized tuple, and $\hat{\bm \beta}\in \mathbb{C}^{ N_b^\text{tx}N_b^\text{rx}N_k^\text{tx}N_k^\text{rx}N_{\omega}\times 1}$ is the sparse path gain vector to estimate. In this way, the   near-field channel can be estimated via compressive sensing schemes such as OMP or CoSaMP theoretically. 

However, a practical problem here is the huge matrix dimension $N_b^\text{tx}N_b^\text{rx}N_k^\text{tx}N_k^\text{rx}N_{\omega}$, especially with the angular resolution $N_b^\text{tx/rx} \gtrsim N_\text{T},\  N_\text{R}\gg 1$, which results to unacceptable computational complexity. Therefore, we herein provide a low-complexity estimation scheme corresponding to this realistic constraint as follows. 

Fortunately, some prior informations are instructive for parameter estimate. One is that the LoS path is unique with strong gain compared with NLoS paths, which means we can coarsely predetermine LoS beamspace sector via conventional mmWave estimate methods \cite{far_field_OMP}. In this way, the LoS codebook can be largely reduced with limited angular sampling points. Assume that the number of remaining directions are $N_{b,\text{sub}}^\text{tx/rx}\ll N_b^\text{tx/rx}$ and then the revised dictionary matrix  $\hat{\bm G}_\text{sub}$ only remains much less columns than $\hat{\bm G}$. Therefore, the LoS component can be rewritten as $\bm y_{\text{L}}=\bm \Psi \hat{\bm G}_\text{sub} \cdot \hat{\bm \beta}_\text{L}$.

On the other hand, the windowing parameters hold a common zero value  $\omega_l=0$ for multiple weak NLoS paths, which means there exists no Tx-Rx coupling relationship for NLoS paths and 2D MMV compressive sensing can be adopted for matrix dimension reduction. Define the polar-domain representation matrix at Tx side as $\bm P^\text{tx}=[\bm b_{N_\text{T}}(b_q^\text{tx},k_q^\text{tx})]\in \mathbb{C}^{N_\text{T}\times N_b^\text{tx}N_k^\text{tx}}$, and similarly $\bm P^\text{rx}$ at Rx side. Define the sensing matrix at Tx/Rx sides as $\bm \Psi^\text{tx}$ and $\bm \Psi^\text{rx}$, respectively ($\bm\Psi = {\bm \Psi^\text{tx}}^T\otimes {\bm\Psi^\text{rx}}^H$). Then the NLoS components can be formulated as $\bm Y_\text{NL}={\bm\Psi^\text{rx}}^H{\bm P^\text{rx}}\bm \Xi {\bm P^\text{tx}}^H\bm\Psi^\text{tx}$, where $\bm \Xi$ is the sparse NLoS gain matrix to estimate. To summarize, the total received signal can be formulated as 
\begin{equation}
	\bm y = \bm \Psi \hat{\bm G}_\text{sub} \cdot \hat{\bm \beta}_\text{L}+\text{vec}({\bm\Psi^\text{rx}}^H{\bm P^\text{rx}}\bm \Xi {\bm P^\text{tx}}^H{\bm\Psi^\text{tx}}).
\end{equation}
 
We herein adopt and modify the greedy-based orthogonal matching pursuit algorithm for the   XL-MIMO. The first is the matching process. In each $i$-th iteration, we define the residual vector as $\bm z^{(i)}$. For LoS component, we calculate the gain as $\bm g^{\text{L},(i)}=(\bm  \Psi\hat{\bm G}_\text{sub})^H\bm z^{(i)}$ and select the most potential gain $g$ and index $j$ with maximum amplitude as $[g_\text{max}^{\text{L},(i)},j^{(i)}]=\max_j |\bm g^{\text{L},(i)}_j|/\|(\bm\Psi\hat{\bm G}_\text{sub})_{:,j}\|_2$. For NLoS component, the gain can be coarsely calculated as $\bm G^{\text{NL},(i)}={\bm P^\text{rx}}^H{\bm\Psi^\text{rx}}\bm Z^{(i)} {\bm \Psi^\text{tx}}^H {\bm P^\text{tx}}$, where $\bm Z^{(i)}=\text{unvec}(z^{(i)})$ denotes the residual matrix. Then we utilize the 2-D OMP and determine the most potential gain and corresponding column/row supporting vector indices as
\begin{equation}
	\begin{small}
[g_\text{max}^{\text{NL},(i)},j^{(i)}_\text{tx},j^{(i)}_\text{rx}]=\max_{m,n} \frac{|\bm G^{\text{NL},(i)}_{m,n}|}{\|({\bm\Psi^\text{rx}}^H\bm P^\text{rx})_{:,m}\|_2 \|({\bm\Psi^\text{tx}}^H\bm P^\text{tx})_{:,n}\|_2}.
\end{small}
\end{equation}

The second is the pursuit procedure. Define the LoS supporting vector indices set is $\bm\Omega^\text{L}$. Similarly $\bm\Omega^\text{NL}_\text{rx}$ and $\bm \Omega_\text{tx}^\text{NL}$ represent the NLoS dominant supporting indices sets at Rx and Tx sides, respectively. All the three sets are initialized as empty and define $\bm z^{(0)}=\bm y$. If the LoS estimate gain is larger, i.e., $g^{\text{L},(i)}_\text{max}>g^{\text{NL},(i)}_\text{max}$, we take the LoS present index $j^{(i)}$ into set $\bm\Omega^\text{L}$ and keep the sets $\bm\Omega_\text{tx}^\text{NL}$ and $\bm\Omega^\text{NL}_\text{rx}$ unchanged. On the contrary if the NLoS estimate gain is larger, i.e., $g^{\text{L},(i)}_\text{max}<g^{\text{NL},(i)}_\text{max}$, we take the NLoS indices $j^{(i)}_\text{tx}, j_\text{rx}^{(i)}$ into $\bm\Omega_\text{tx}^\text{NL}$ and $\bm\Omega^\text{NL}_\text{rx}$. Finally, we update the gain estimate as $\bm \beta^{(i)} = (\bm\Psi \bm G^{(i)})^\dagger \bm y$, where $\bm G^{(i)}=[(\hat{\bm G}_\text{sub})_{:,\bm\Omega^\text{L}}, (\bm P^\text{tx,*})_{:,\bm\Omega^\text{NL}_\text{tx}}\otimes (\bm P^\text{rx})_{:,\bm\Omega^\text{NL}_\text{rx}})]$ is the temporary dictionary matrix and $\bm \beta^{(i)}$ is the total gain estimate including both LoS and NLoS components. 

Finally, the residual vector is calculated as 
\begin{equation}
	\bm z^{(i+1)} = \bm y - (\bm \Psi\bm G^{(i)}) (\bm\Psi\bm G^{(i)})^\dagger \bm y,
\end{equation}
and the procedures are finished inside one iteration for the unified LoS/NLoS   near-field XL-MIMO estimation scheme. The detailed expression is shown in Algorithm 1 as follows.

\begin{algorithm}[htb] 
	\normalem
	\caption{Unified LoS/NLoS OMP  estimation scheme for   XL-MIMO (XL-UOMP Algorithm)} 
	\label{alg1} 
	\begin{algorithmic}[1] 
		\REQUIRE Sensing matrix $\bm \Psi^\text{tx}\in \mathbb{C}^{N_\text{T}\times T}$ and $\bm\Psi^\text{rx}\in \mathbb{C}^{N_\text{R}\times T}$ at Tx/Rx sides, Received signal $\bm y$ along several pilot slots, Pre-determined LoS dictionary matrix $\hat{\bm G}_\text{sub}$, Maximum iteration number $N_\text{iter}$. 
		
		\ENSURE channel matrix $\bm H^\text{est}$
		
		\STATE \textbf{Initialize:} Residual vector $\bm z^{(0)}=\bm y$, $\bm Z=\text{unvec}(\bm z)$, sets $\bm\Omega^\text{L}=\bm\Omega^\text{NL}_\text{tx}=\bm\Omega_\text{rx}^\text{NL}=\bm\varnothing$
		
		\FOR{Iteration number $i=0$ to $N_\text{iter}$}
		
	 \leftline{\ \ \ \ \ \textit{\% LoS Component Matching}}
		
		\STATE $\bm g^{\text{L},(i)}\leftarrow(\bm  \Psi\hat{\bm G}_\text{sub})^H\bm z^{(i)}$
		
		\STATE $[g_\text{max}^{\text{L},(i)},j^{(i)}]\leftarrow \max_j |\bm g^{\text{L},(i)}_j|/\|(\bm\Psi\hat{\bm G}_\text{sub})_{:,j}\|_2$
		
	 \leftline{\ \ \ \ \   \textit{\% NLoS Component Matching}}
		
		\STATE $\bm G^{\text{NL},(i)}\leftarrow {\bm P^\text{rx}}^H{\bm\Psi^\text{rx}}^H\bm Z^{(i)} {\bm \Psi^\text{tx}} {\bm P^\text{tx}}$
		
		\STATE $[g_\text{max}^{\text{NL},(i)},\{j^{(i)}_\text{tx},j^{(i)}_\text{rx}\}]\leftarrow {\displaystyle \max_{m,n}} \frac{|\bm G^{\text{NL},(i)}_{m,n}|}{\|(\bm\Psi^\text{rx}\bm P^\text{rx})_{:,m}\|_2 \|(\bm\Psi^\text{tx}\bm P^\text{tx})_{:,n}\|_2}$ 
		
	 \leftline{\ \ \ \ \  \textit{\% Comparision and Pursuit}}
		
		\IF{$g^{\text{L},(i)}_\text{max}\geq g^{\text{NL},(i)}_\text{max}$}
		\STATE $\bm\Omega^\text{L}\leftarrow \bm\Omega^\text{L}\cup \{j^{(i)}\}$
		\ELSE
		\STATE $\bm\Omega^\text{NL}_\text{tx}\leftarrow \bm\Omega^\text{NL}_\text{tx}\cup \{j_\text{tx}^{(i)}\}$, $\bm\Omega^\text{NL}_\text{rx}\leftarrow \bm\Omega^\text{NL}_\text{rx}\cup \{j_\text{rx}^{(i)}\}$
		\ENDIF
		
		 \leftline{\ \ \ \ \   \textit{\% Reconstruction and Residual Update} }
		
		\STATE $\bm G^{(i)}\leftarrow [(\hat{\bm G}_\text{sub})_{:,\bm\Omega^\text{L}}\  ,\  (\bm P^\text{tx,*})_{:,\bm\Omega^\text{NL}_\text{tx}}\otimes (\bm P^\text{rx})_{:,\bm\Omega^\text{NL}_\text{rx}})]$
		
		\STATE $\bm \beta^{(i)} \leftarrow (\bm\Psi \bm G^{(i)})^\dagger \bm y$
		
		\STATE $	\bm z^{(i+1)} \leftarrow \bm y - (\bm \Psi\bm G^{(i)}) (\bm\Psi\bm G^{(i)})^\dagger \bm y$\\ $\bm Z^{(i+1)}\leftarrow \text{unvec}(\bm z^{(i+1)})$
		
		\ENDFOR
		
		\STATE $\bm H^\text{est}\leftarrow \text{unvec}(\bm G^{(i)}\bm \beta^{(i)})$
		
	\end{algorithmic}
\end{algorithm}

As for algorithmic complexity, the dominant computation locates in the matching and reconstruction parts. For brevity, we define the dimension of $\hat{\bm G}_\text{sub}$ as $N_\text{T}N_\text{R}\times N_\text{sub}$. The LoS/NLoS matching contains several matrix multiplications and is evaluated  approximately as $\mathcal{O}(T^2N_\text{T}N_\text{R}N_\text{sub})$. For the reconstruction part in $i$-th iteration, the main complexity results from the Moore-penrose inverse, which can be approximated as $\mathcal{O}(T^2N_\text{T}N_\text{R}i)$. Therefore, the total complexity after $N_\text{iter}$  iterations can be formulated as $\mathcal{O}(T^2N_\text{T}N_\text{R}N_\text{iter}(N_\text{sub}+N_\text{iter}))$, which is quite reduced compared with direct OMP via (\ref{formula_OMP}). It should be noted that the framework and estimation can be similarly extended to uniform planar array (UPA), RIS \cite{RIS_estimation} and wideband near-field communications such as OFDM or OTFS-based scenarios \cite{otfs}.  
   
\section{Simulation Results}

The parameter setting is as follows. Both transmitter and receiver are equipped with ULA $N_\text{T}=N_\text{R}=256$, RF chains $N_\text{RF}=5$ and carrier frequency $f_c=100 \text{GHz}$. The total channel contains one LoS component and three NLoS paths, while the nLoS paths have $-25\sim -20 \text{dB}$ power loss compared to the LoS one.  LoS/NLoS AoAs and AoDs are randomly generated following $\mathcal{U}(-\frac{\pi}{3},\frac{\pi}{3})$ and the LoS Tx/Rx tilt angle $\phi_\text{rx}^\text{L}$ follows $\mathcal{U}(-\frac{\pi}{2},\frac{\pi}{2})$, which is similar to \cite{luyu}. The Tx/Rx central distance is set as $35\text{m}$ inside the Fresnel near-field region. For convenience, we define the pilot length $T$ as the dimension of precoding/combining matrices $\bm \Psi^\text{tx}\in \mathbb{C}^{N_\text{T}\times T}, \bm \Psi^\text{rx}\in \mathbb{C}^{N_\text{R}\times T}$, where all elements are randomly selected from $\{-1,1\}$ before normalization. We herein compare with two benchmarks, i.e., the conventional far-field DFT codebook and the near-field polar-domain representation (without consideration of the Tx/Rx coupled coefficient). For all estimation schemes, the maximum potential path number to estimate, i.e., the maximum iteration number $N_\text{iter}$ during OMP, is set as $8$.

\begin{figure}[!t]
	\centering
	\includegraphics[width=0.7\linewidth]{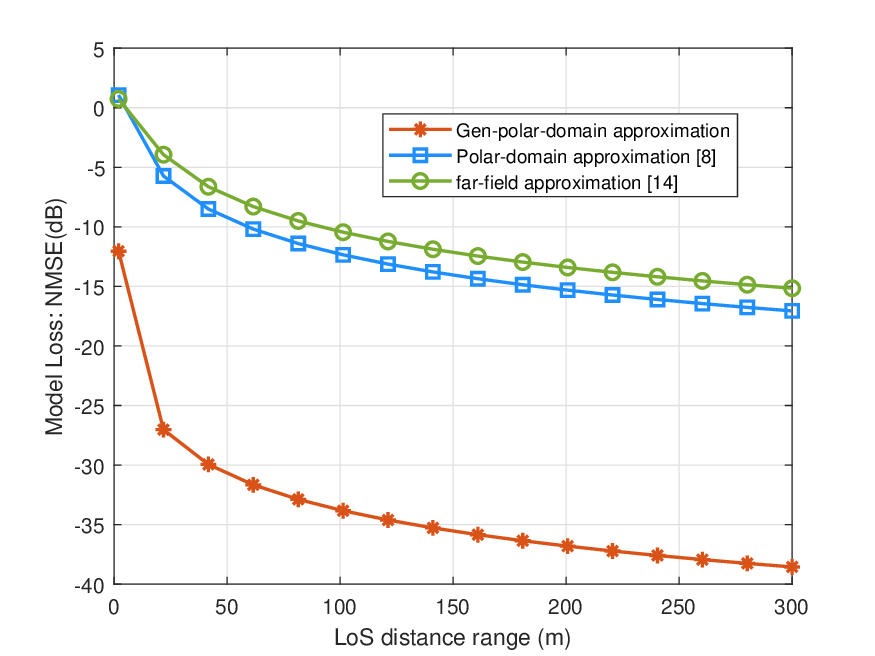}
	\caption{Model loss for the second-order Taylor expansion-based generalized-polar-domain approximation.}
	\label{model_loss}
\end{figure}

First we compare the LoS model approximation error for several modelling schemes, i.e., the conventional far-field approximation, polar-domain representation and the proposed generalized-polar-domain formulation (\ref{distance_formulation}) in Fig. \ref{model_loss}. We can clearly observe that the proposed second-order Taylor-expansion-based scheme can obtain the lowest modelling approximation error under the perfect LoS path parameter information.

\begin{figure}[!t]
	\centering
	\begin{minipage}[t]{0.5\linewidth}
		\centering
		\includegraphics[width=1\linewidth]{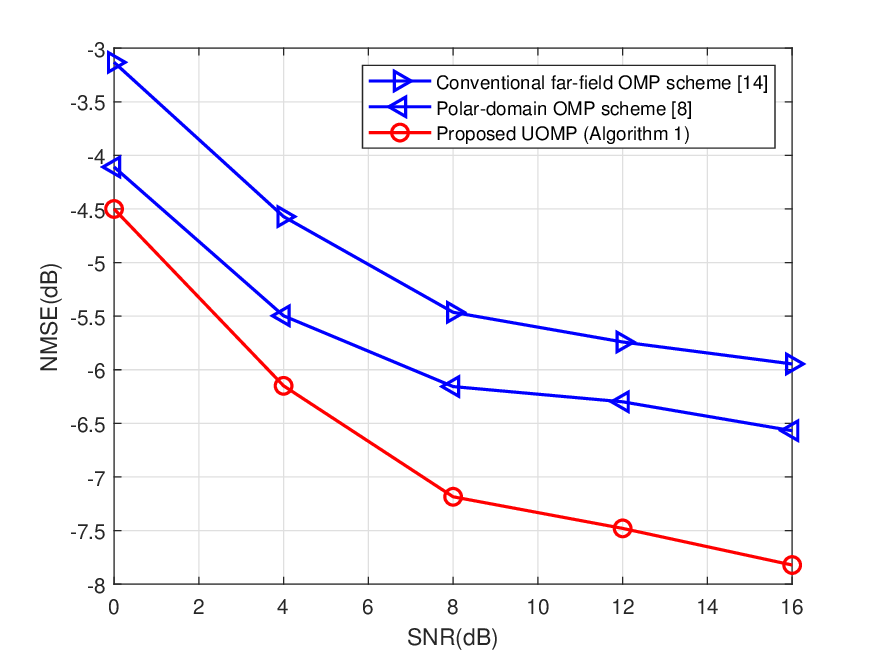}
	\end{minipage}%
	\begin{minipage}[t]{0.5\linewidth}
		\centering
		\includegraphics[width=1\linewidth]{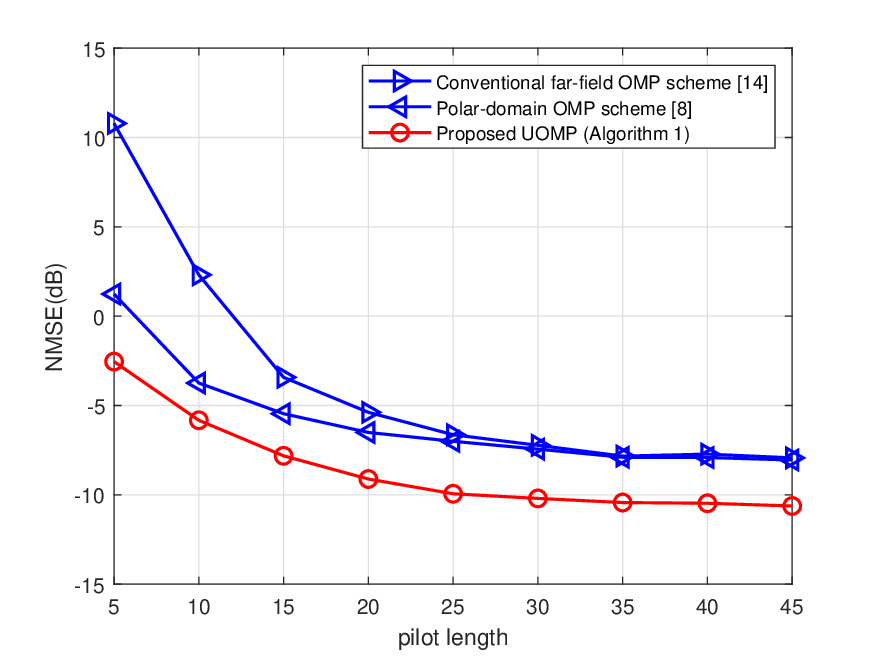}
	\end{minipage}
	\caption{NMSE against SNR and pilot length: (a) with pilot length fixed as 25; (b) with SNR fixed as $20 \text{dB}$.}
	\label{NMSE_SNR}
\end{figure}

Then we elaborate the normalized mean square error (NMSE) verse signal-to-noise ratio (SNR) with fixed pilot length $T=25$ in Fig. \ref{NMSE_SNR}. Since the additional Tx/Rx coupled coefficient $\omega_0$ is considered in our proposed XL-UOMP scheme, the estimation performance is largely enhanced via our proposed UOMP scheme for  XL-MIMO system. It should be noted that we only consider the on-grid estimation schemes for brevity, where the beam mismatch causes the power leakage problem and limits the accuracy improvement at high-SNR scenario. Fig. \ref{NMSE_SNR} also presents the estimate NMSE verse pilot length $T$, with SNR fixed to $20\  \text{dB}$. Notice that the gap between our proposed XL-UOMP scheme and conventional methods always holds, regardless of the amount of estimate overhead from $T=10$ to $T=45$. This benefits from the accurate LoS channel formulation with the windowing matrix $\bm V_0$ in (\ref{Van_LoS}). 

\section{Conclusion}
In this paper, we provide one concise closed-form channel formulation for   XL-MIMO. All the distinctions between LoS and NLoS paths, spherical and planar-wavefront propagations, single and multiple-antenna user scenarios are all unified under this closed-form expression. We further propose one low-complexity XL-UOMP algorithm for XL-MIMO channel estimation. The simulation results demonstrate the superiority of the proposed algorithm on both estimation accuracy and pilot consumption.

\appendix
\begin{IEEEproof}
	Following the expression in (\ref{LoS_model}), we can reformulate the LoS channel matrix as
	\begin{equation}
	\bm H^\text{L}=\beta_0^\text{L}\text{diag}(\bm b_{N_\text{R},0})\cdot\bm V_0\cdot \text{diag}(\bm b_{N_\text{T},0}^H),
	\end{equation} 
	where we can easily find that the LoS channel rank is exactly the same with the rank of windowing matrix $\bm V_0$. In another word, the LoS pipeline number is determined by the windowing parameter $\omega_0$. Although the windowing matrix $\bm V_0$ contains \emph{full rank of $\min\{N_\text{T},N_\text{R}\}$} according to Vandermonde property, \emph{the condition number $\frac{\lambda_\text{max}(\bm V_0)}{\lambda_\text{min}(\bm V_0)}$} is quite large which indicates that the number of supportive eigenvalues is limited.

	Based on the Parseval's Theorem, the total power of matrix $\bm V_0$ is formulated as 
	\begin{equation}
	P_\text{total}=\|{\bm F^\text{rx}}^H\bm V_0\|_F^2=\|\bm  V_0\|_F^2=N_\text{T}N_\text{R}.
	\label{P_total}
	\end{equation}
	And we observe that the dominant power of $\bm V_0(\omega_0)$ is mainly located inside the subspace spanned by the former $\omega_0N_\text{R}N_\text{T}$ Fourier bases, which can be obtained via 
	\begin{equation}
	\begin{small}
	\arraycolsep=1.0pt\def\arraystretch{1.5}
	\begin{array}{cll}
	P_\text{sub}&=&\bigg\|\big[\bm f_{-\frac{N_\text{R}}{2}+1}^\text{rx},\dots,\bm f_{\omega_0N_\text{R}N_\text{T}-\frac{N_\text{R}}{2}}^\text{rx}\big]^H\bm V_0\bigg\|_F^2\\
	&=&\displaystyle \sum_{i=-\frac{N_\text{R}}{2}+1}^{\omega_0N_\text{R}N_\text{T}-\frac{N_\text{R}}{2}}\sum_{n_1=0}^{N_\text{T}-1}\|\bm v_{n_1}^H\bm f_i^\text{rx}\|^2\\
	&=&\displaystyle\frac{1}{{N_\text{R}}} \sum_{i=0}^{\omega_0N_\text{R}N_\text{T}-1}\sum_{n_1=0}^{N_\text{T}-1} \frac{\sin^2\big(\pi (\omega_0 N_\text{R}n_1-i)\big)}{\sin^2 \big(\frac{\pi}{N_\text{R}} (\omega_0 N_\text{R}n_1-i)\big)}\\
	&\overset{(a)}{\approx}& \displaystyle\frac{1}{{N_\text{R}}} \sum_{i=0}^{\omega_0N_\text{R}N_\text{T}-1} \int_{-1}^{\omega_0N_\text{R}N_\text{T}}\frac{\sin^2\big(\pi (x-i)\big)}{\sin^2 \big(\frac{\pi}{N_\text{R}} (x-i)\big)}\text{d}\frac{x}{\omega_0 N_\text{R}}\\
	&\overset{(b)}{>}&\displaystyle\frac{1}{\omega_0 N_\text{R}^2} \sum_{i=0}^{\omega_0N_\text{R}N_\text{T}-1}\int_{i-1}^{i+1}\frac{\sin^2\big(\pi (x-i)\big)}{\sin^2 \big(\frac{\pi}{N_\text{R}} (x-i)\big)}\text{d} x\\
	&=&\displaystyle \frac{1}{\omega_0 N_\text{R}^2} \omega_0N_\text{R}N_\text{T}\int_{-1}^{1}\frac{\sin^2\big(\pi x\big)}{\sin^2 \big(\frac{\pi}{N_\text{R}} x\big)}\text{d}x\\
	&\overset{(b)}{\approx}&\displaystyle 0.9028 N_\text{T}N_\text{R},
	\end{array}
	\label{P_subspace}
	\end{small}
	\end{equation}
	where columns of windowing matrix are defined as $[\bm v_0,\dots,\bm v_{N_\text{T}-1}]=\bm V_0$. $(a)$ is approximated by replacing the summation with integral and $(b)$ is derived by only remaining the mainlobe power of Dirichlet function $\frac{\sin Na}{\sin a}$. From (\ref{P_total}) and (\ref{P_subspace}) we can get that over $90\%$ powers are gathered in this subspace, which contains a space dimension of $\omega_0N_\text{T}N_\text{R}$. The same conclusion can be deduced by $\bm F^\text{tx}$ at Tx side. This completes the proof of Theorem 1.
\end{IEEEproof}

\bibliographystyle{ieeetr}
\normalem
\bibliography{bare_jrnl.bib}

\ifCLASSOPTIONcaptionsoff
  \newpage
\fi

\end{document}